\documentclass[11pt]{article}
\usepackage{newpasp,epsf}
\def\edcomment#1{\iffalse\marginpar{\raggedright\sl#1\/}\else\relax\fi}
\reversemarginpar
\newcommand{\gtsim}{\mbox
{{\raisebox{-0.4ex}{$\stackrel{>}{{\scriptstyle\sim}}$}}}}
\newcommand{\ltsim}{\mbox
{{\raisebox{-0.4ex}{$\stackrel{<}{{\scriptstyle\sim}}$}}}}

\begin{document}
\title{Cosmological Studies from Radio Source Samples} 
 \author{Steve Rawlings}
\affil{University of Oxford, Astrophysics, Keble Road, Oxford, OX1 3RH, UK}

\begin{abstract}
I review some recent cosmological studies based
on redshift surveys of radio sources selected at low
frequencies. The accretion rate onto the central black hole
is identified as the basis of a crude physical
division of the low-frequency population into two sub-populations:
the first comprises `Eddington-tuned' 
(high accretion rate) quasars and their torus-hidden counterparts; the 
second comprises `starved quasars' like M87.
There exist remarkable similarities between the shapes and 
evolutionary behaviours of the 
luminosity functions of radio sources and radio-quiet quasars;
all luminous AGN are $\sim300-$times rarer at 
epochs corresponding to $z \sim 0$ than at $z \sim 2.5$.
I argue that any evidence that quasars
were intrinsically rarer at $z \sim 5$ than at $z \sim 2.5$ 
is as yet both tentative and indirect. 
A simple calculation suggests that the 
radio source population has been over-looked as a potentially
important contributor to the entropy budget of the Universe.
A recent sub-mm survey of radio sources is used to
demonstrate a connection between the events which 
trigger jets and intense bursts of star formation, and a close
link between the histories of star formation and 
AGN activity is proposed.
I discuss the aims and methods of future large redshift
surveys of radio sources, emphasising the importance
of dovetailing these with the development of
robust physical models for radio sources and
their epoch-dependent environments.
\end{abstract}

\section{Redshift surveys of low-frequency radio sources}
\label{sec:intro}

We were all excited at this meeting by the 
prospect of the GMRT pushing low-frequency astronomy to
sensitivities significantly below those of the previous generation of 
surveys: e.g. the $\sim5 \sigma$ detection levels 
of 151\,MHz flux density $S_{151} \sim 100 ~ \rm mJy$
reached by the Cambridge 7C survey (e.g. McGilchrist et al.\ 1990). 
Nevertheless,
the next generation of low-frequency 
continuum surveys will
continue to be dominated by extragalactic sources whose  
large radio luminosities come ultimately from 
energy-releasing processes associated 
with super-massive black holes. In this sense most of the sources in a 
GMRT survey will not be so different from those in the 
3CRR survey and, as we have recently been reminded
by Longair (1999), these sources have been studied
for nearly forty years. 
So, what new things can we learn from low-frequency radio surveys deeper than
3CRR? As a partial answer to this question I will develop
two themes illustrated by the 
increased coverage of the 151\,MHz luminosity\footnotemark $L_{151}$,
redshift $z$ plane (Fig.~1) provided by recently completed
redshift surveys of radio sources. 

\begin{figure}[!h]
\label{fig:pz}
\plotone{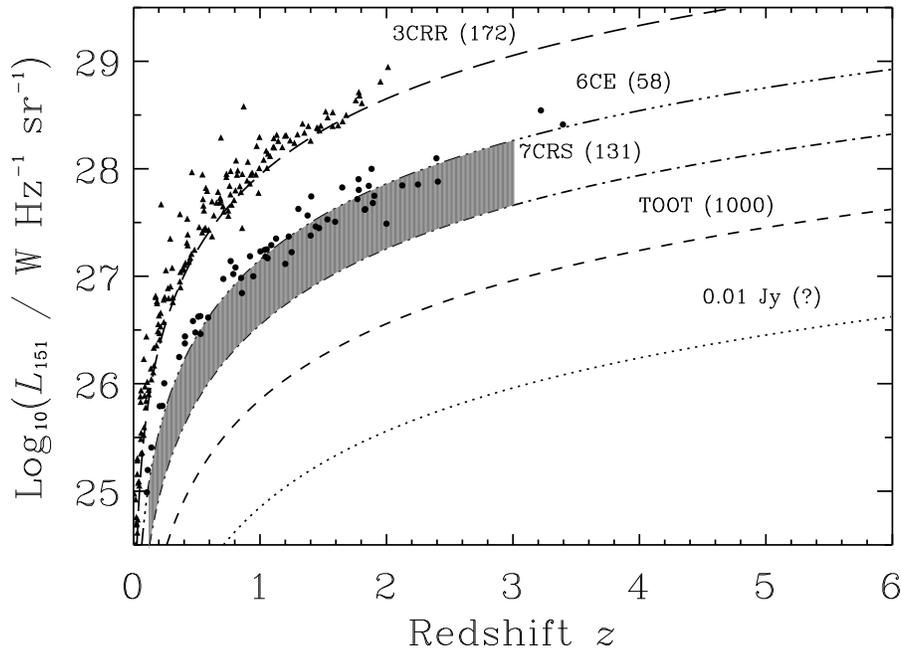}
\caption{The coverage of the 151\, MHz luminosity $L_{151}$ 
versus redshift $z$ plane for existing, planned and possible future 
redshift surveys of low-frequency selected radio sources. The number of
objects in each survey is given in brackets, slightly above the
locus of the relevant 
$S_{151}$ lower limit
(taking $\alpha = 0.8$). References for the named 
surveys are: 3CRR (triangles) --
Laing, Riley \& Longair 
(1983); 6CE (circles) -- Rawlings, Eales \& Lacy (2000);
7CRS (shaded region) -- Willott et al.\ (2000d),
Lacy et al.\ (1999), Blundell et al.\ (2000); {\sc TOOT} --
Sec.~6.
}
\end{figure}

\footnotetext{Throughout this paper I take
$H_{\circ}=50~ {\rm km~s^{-1}Mpc^{-1}}$, $\Omega_{\rm M}=0.3$ and
$\Omega_ {\Lambda}=0.7$; radio spectral index, $\alpha$, is
defined by $S_{\nu} \propto \nu^{-\alpha}$, where $S_{\nu}$ is the
flux-density at frequency $\nu$; and radio and optical 
narrow-line/bolometric
luminosities are measured in units of $\rm W ~ Hz^{-1} ~ sr^{-1}$
and W respectively.
}

The first theme concerns the demography of the low-frequency radio population.
The tight distribution of the 3CRR points on Fig.~1
results from the conspiracy of
three features of the radio luminosity function (RLF) ---
the density 
$\rho_{\rm R} \equiv [{\rm d} \rho(L_{151}, z) / {\rm d} \log_{10} L_{151}]$
of radio sources per unit
co-moving volume and per unit logarithm of radio luminosity ---
namely: (i) the points hug the locus of the 3CRR
$S_{151}$ limit because at any given
$z$, the RLF is a steeply declining function of $L_{151}$; and (ii)
the points are fairly uniformally distributed along this locus
because out to $z \sim 2$ both the normalisation of the RLF 
and the observable co-moving volume are steeply increasing functions of $z$.
A corollary of these features is that the RLF is well determined 
close to the locus of the $S_{151}$ limit but estimates 
elsewhere in the $L_{151}$--$z$ plane require
extrapolations of what is clearly a rapidly varying function 
(although data like the low-frequency radio source counts provide
additional constraints). 
Constraining the RLF at regions of the $L_{151}$--$z$
plane corresponding to higher {\em aerial} densities of radio sources
(e.g. the 7C Redshift Survey, 7CRS,
region of Fig.~1) is essential for many studies which
utilise radio sources as cosmological probes: e.g.\ gravitational
lensing studies which constrain cosmic geometry (e.g.\ Kochanek 1996) and
clustering studies which map out large-scale structure
(e.g.\ Magliocchetti et al.\ 1998). 

Of more direct concern to us here is the ability to understand the RLF
in terms of physical models for radio sources and their cosmic evolution.
Ignoring for now any variations in radio source environment, what we would
really like to measure is the density
$\rho_{\rm P} (Q,t,z)$, where $Q$ is the jet power
(Scheuer 1974; Rawlings \& Saunders 1991), and $t$ is the source age, the 
time after the jet-triggering event. From $\rho_{\rm P} (Q,t,z)$ it is possible
to calculate physically meaningful quantities like
the birth function -- the number of jets triggered per co-moving 
volume per unit cosmic time --- and the total entropy injected 
into the intracluster medium (ICM) by the radio source population.
Successful mapping between the `physical' density $\rho_{\rm P}
(Q,t,z)$ and the 
`observable' density $\rho_{\rm O} (L_{151},D,z)$ 
(and hence, by marginalising the projected radio source size $D$, 
$\rho_{\rm R}$) requires both a good physical model and samples providing 
adequate coverage of Fig.~1. Inadequate 
coverage leads to unbreakable degeneracies 
between the key parameters: 
Blundell \& Rawlings (1999) have recently 
emphasised one such effect which means that the known highest-$z$
radio galaxies are biased towards objects with
extremely high $Q$ and relatively small $t$, some complicating effects 
of which will be discussed in Sec.~5.

This introduces the second important theme illustrated by Fig.~1 --
the use of radio sources as probes of the very high redshift  
Universe. The highest redshift object in the 3CRR sample is at
$z \approx 2$. The lack of higher-$z$ 3CRR objects is not due
to an abrupt high-$z$ cut-off in the RLF, but due instead to the
intrinsic rarity of radio sources with the extreme radio luminosities
[$\log_{10} (L_{151}) \gtsim 29$] required to exceed the $S_{151}$
limit of 3CRR at high redshift.
At a given redshift the finite (co-moving) volume 
on our light cone depends only on the assumed cosmological model, 
and since the 3CRR survey covers a good fraction of the entire sky, 
the only way of finding significant numbers of 
higher-$z$ radio sources is by studying objects with 
luminosities similar to those of the 3CRR objects. This requires 
spectroscopic surveys of fainter radio-selected 
samples; the Oxford-led 6CE and 7CRS redshift surveys 
(see Fig.~1) have demonstrated that one can achieve
close to $100 \%$ redshift completenesses for such surveys
using 4-m class optical telescopes. This has
allowed the construction of sizeable and complete samples of $z > 2$ 
radio sources from follow-up of fairly small sky areas.

\section{Sub-populations of low-frequency radio sources}
\label{sec:subpop}

The traditional way of dividing the low-frequency radio population
into two distinct sub-populations is via the Fanaroff \& Riley (1974)
scheme based on radio structure: 
crudely, FRII galaxies have jet-terminating
compact hotspots at their outer extremities, whilst FRI galaxies do not.
Fanaroff \& Riley showed that this FRI/FRII structural division 
corresponds closely to a division in radio luminosity at
$\log_{10} (L_{151}) \approx 25.5$ although more recent studies 
(Ledlow \& Owen 1996) have suggested an additional
dependence on host galaxy luminosity. However, following 
Hine \& Longair (1979) and Laing et al.\ (1994), 
there is another, arguably more fundamental,
way of dividing the population into two classes.
I will adopt two underlying assumptions: (i) that all radio galaxies harbour
super-massive ($\sim10^{9} ~ \rm M_{\odot}$) black holes with a 
low spread in black hole mass (see Willott et al.\ 1999), and 
(ii) that higher accretion rates onto the black hole boosts:
(a) the quasar optical luminosity, and hence the 
narrow emission line luminosity; and (b) the jet power, and hence the
radio luminosity (Rawlings \& Saunders 1991; Serjeant et al.\ 1998).
I will neglect the effects of 
scatter in correlations between $L_{151}$ and, for example, 
the [OII] emission line luminosity $L_{\mathrm [OII]}$
(e.g. Baum \& Heckman 1989; Willott et al.\ 1999).

Above $\log_{10}(L_{\mathrm [OII]}) \approx 35$ 
[or $\log_{10}(L_{151}) \approx 26.5$], the
`quasar fraction' -- the fraction of objects with observed
broad emission lines -- is $\approx 0.4$ regardless of luminosity
or redshift, a result which is consistent with a simple
unified scheme in which each object has an
obscuring torus with a half-opening angle $\theta_{\rm trans} 
\approx 53^{\circ}$ (Willott et al.\ 2000b). In other words this 
appears to be a homogeneous `Eddington-tuned'
sub-population comprising objects
whose black holes are accreting at some sizeable and roughly 
fixed fraction of their Eddington-limited rates, and whose nuclei have
well-developed obscuring tori.

Below $\log_{10}(L_{\mathrm [OII]}) \approx 35$
[or $\log_{10}(L_{151}) \approx 26.5$], the quasar fraction drops 
abruptly, and a second sub-population dominates. 
The archetype of this sub-population is the
nearby FRI radio galaxy M87 which seems certain
to contain a super-massive black hole accreting at 
a tiny ($< 10^{-4}$) fraction of its Eddington-limited 
rate (e.g. Ford et al.\ 1994).
Chiaberge, Capetti \& Celotti (1999) discuss 
how the seemingly clear view of optical
synchrotron from the base of the M87 jet
argues against the existence of an obscuring torus in this
radio galaxy.

Because the break between these `starved quasar'
sources and the Eddington-tuned population lies about
one dex higher in $L_{151}$ than the FRI/FRII division, there are
starved quasars with both FRI and FRII radio structure, whereas
the Eddington-tuned sources are almost exclusively FRIIs.
The existence of FRIIs lacking an intrinsically bright quasar 
nucleus is supported by cases in which narrow emission lines
are either weak or absent (Laing et al.\ 1994), and, as for M87,
the detection of optical synchrotron from some FRII nuclei rules out 
some types of obscuring tori in some FRII objects 
(Chiaberge, Capetti \& Celotti 2000).

Being based on accretion rate, the division proposed here is much  
more intimately tied to physical processes near the
black hole than the FRI/FRII divide
which is likely to be a larger-scale
magnetohydrodynamical effect driven by the 
dependence of the stability of jets on jet power and environment 
(e.g. Kaiser \& Alexander 1997). Since Eddington-tuned
radio sources always contain powerful, and hence stable, jets 
it is no surprise that they almost exclusively develop FRII radio structures. 
The less powerful jets in `starved quasars' can produce either FRI or FRII
radio sources depending on their precise jet power and environment.

The accretion rate based division proposed here is undoubtedly
too simplistic: as suggested by Rawlings \& Saunders (1991),
variations in accretion rate from object-to-object
might be a far smoother affair than a clean dichotomy. Indeed, 
the possibility that the low quasar fraction at low luminosities 
is due to lower values of $\theta_{\rm trans}$ for weaker
quasar nuclei,
rather than the intrinsic lack of such nuclei,
has received support from the recent evidence for 
accretion-disc excited broad lines in the pole-on low-luminosity radio
source BL Lacertae (Corbett et al.\ 2000). However, as as we shall see in 
Sec.~3, the two sub-population hypothesis proposed here
provides a simple luminosity-dependent density evolution (LDDE)
model on which to base our first stabs at mapping and
understanding the cosmic evolution of the radio source population.

\section{Cosmic evolution between $z \sim 0$ and $z \sim 2.5$}
\label{sec:lowz}

\begin{figure}[!h]
\label{fig:evol}
\plottwo{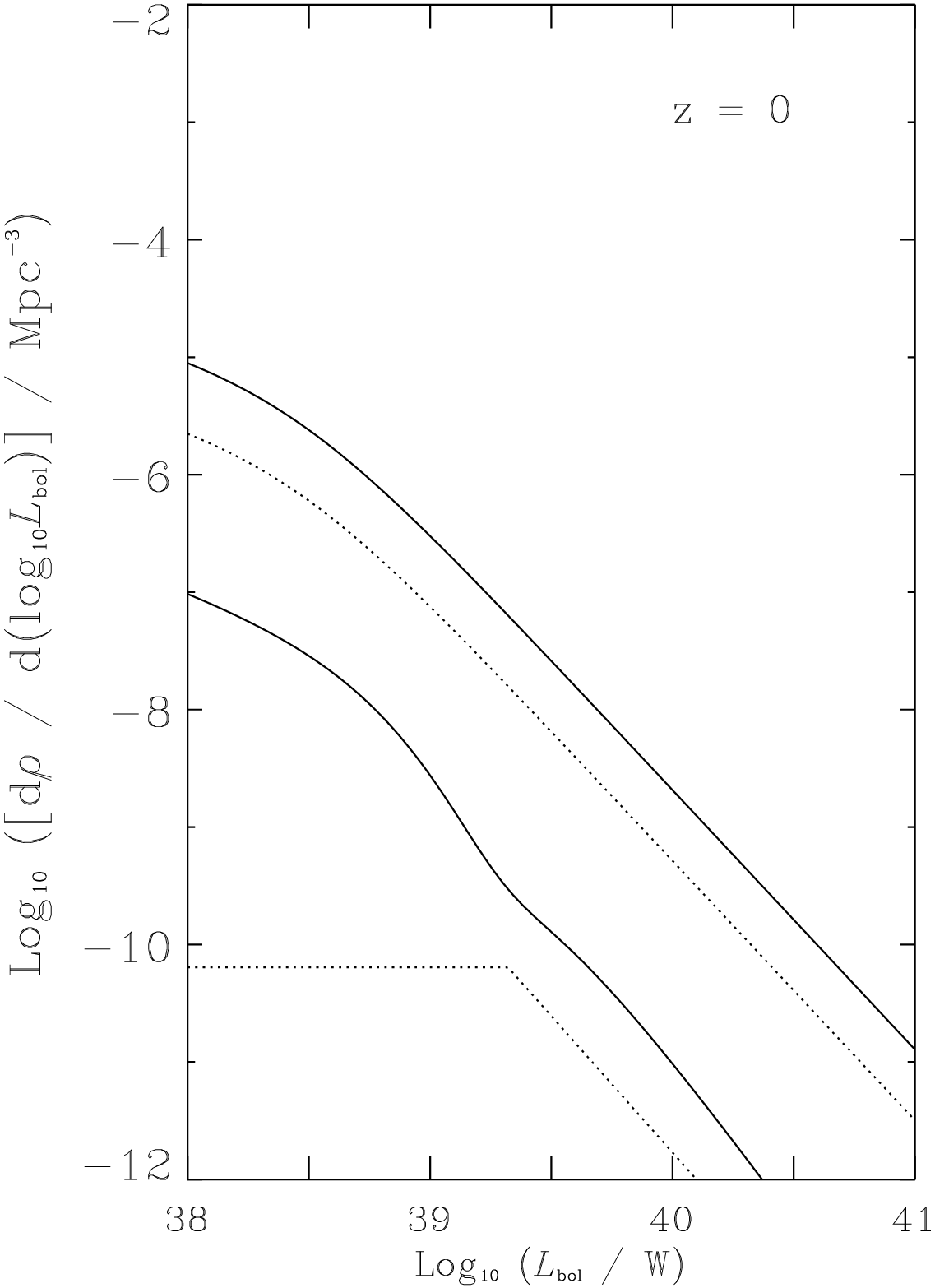}{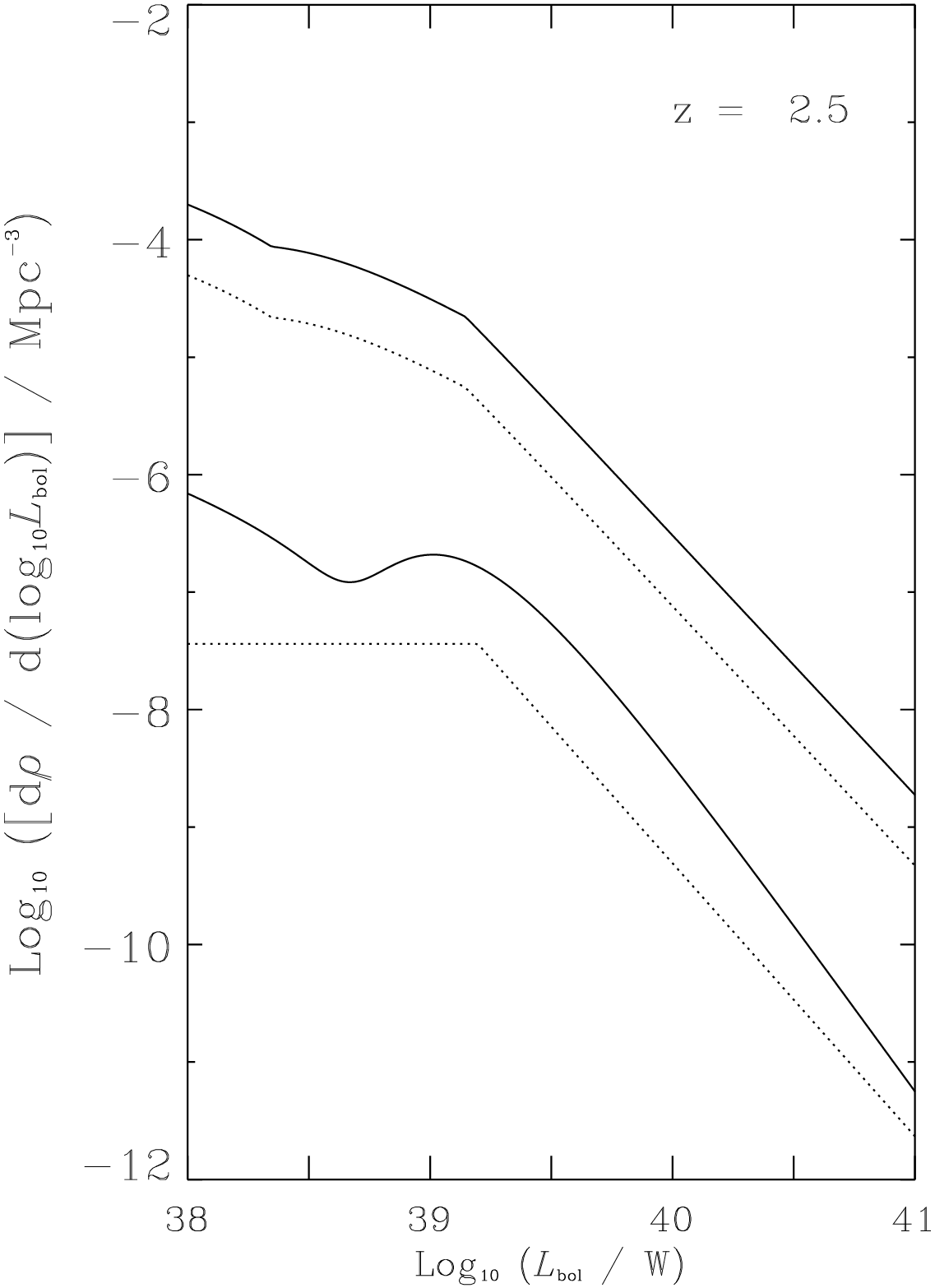}
\caption{Comparison of the luminosity functions of 
active galaxies selected at X-ray and radio
wavelengths at $z = 0$ ({\bf left}), and $z=2.5$ ({\bf right}).
The upper solid line is a simple scaling of the
LDDE2 model of Miyaji et al.\ (2000) for the
soft X-ray luminosity function (upper dotted line), adopting 
a scaling factor $\sim4$ to correct for the large
population of Compton-thin, photoelectrically-absorbed quasars 
required to provide the X-ray background (e.g.\ Wilman \& Fabian 1999); 
I have assumed that $\approx 2$ per cent of 
the bolometric luminosity $L_{\rm bol}$ of a naked quasar emerges in the  
$0.5-2$ keV band (Elvis et al.\ 1994).
The lower solid line shows the model steep-spectrum 
RLF from Willott et al.\ (2000c), and the lower dashed line is a model RLF
for just the steep-spectrum quasars from Willott et al.\ (1998); 
$L_{151}$ and ${\rm d} \rho / {\rm d} \log_{10} L_{151}$
have been converted to $L_{\rm bol}$ and 
${\rm d} \rho / {\rm d} \log_{10} L_{bol}$ respectively, assuming
that $L_{\rm bol} =  10^{16.2} \times L_{151}^{0.85}$ 
(Willott et al.\ 1999). 
}
\end{figure}

Willott et al.\ (2000c) have used the 3CRR, 6CE and 7CRS datasets
illustrated in Fig.~1 (alongside other pertinent
data like the low-frequency radio source counts)
to constrain parametric models for the 
cosmic evolution of the RLF; these models were built on the assumption that 
the two sub-populations described in Sec.~2 
undergo differential density evolution. 
They found that the integrated
co-moving number density $\Upsilon = \int [{\rm d} \rho / 
d \log_{10} L_{151}] ~d \log_{10} L_{151}$ of the
Eddington-tuned sources rises by about 2.5 dex between $z \sim 0$ 
and $z \sim 2.5$, and that the $\Upsilon$ of the starved quasar 
sub-population rises by about 1 dex between $z \sim 0$ and $z \sim 1$.
The superposition of these sub-populations produces an overall
steep-spectrum RLF in fair agreement with a previous determination 
based on high-frequency (2.7\,GHz) selected samples
(Dunlop \& Peacock 1990).
The `break luminosity' in the overall RLF is the
swap-over point between the two sub-populations. Because these
sub-populations undergo differential density evolution, the break
point moves with redshift in a way which 
is similar to the `pure luminosity evolution' (PLE) model 
fitted by Dunlop \& Peacock to their data. 
The Willott et al.\ LDDE model is more attractive than a PLE
model because it is hard to see how the latter can provide a 
physical description of evolution given that the
ages of FRII radio sources are known (e.g. Blundell \& Rawlings 2000) 
to be a tiny fraction of the Hubble time.

I have chosen to illustrate these results in a way 
(Fig.~2) which emphasises some remarkable
similarities between the evolution of the radio population, and 
quasars in general. First, the shapes of the radio and
X-ray luminosity functions are almost identical, and it is
interesting in this regard
that the most recent estimates of the X-ray luminosity 
function (Miyaji, Hasinger \& Schmidt 2000) prefer LDDE to PLE
models. Second, the evolutions with redshift are very similar.
We may soon become persuaded of a third similarity
related to the luminosity-dependent obscuration of the
quasar population. Studies of the RLF (Willott et al.\ 2000b) show that the 
fraction of objects with observed broad lines rises from $\ltsim 0.1$ at 
$\log_{10} (L_{bol}) \sim 39$ (the Eddington-limit of a 
$\sim10^{8} ~ \rm M_{\odot}$ black hole) to $\approx 0.4$ 
at high luminosities. For the X-ray population at 
$\log_{10} (L_{bol}) \sim 39$, the fraction of naked
quasars can only be $\ltsim 0.2$ since the hard X-ray background
is dominated by a photoelectrically-absorbed, but
Compton-thin population (e.g.\ Wilman \& Fabian 1999). Follow-up of
{\sc Chandra} surveys should confirm 
whether or not the fraction of naked quasars rises with luminosity, 
as is the case for the radio population, and
as is predicted by the popular `receding-torus' model (Lawrence 1991).
First results from {\sc Chandra} (e.g.\ Mushotzky et al.\ 2000;
Crawford et al.\ 2000) suggest
that it may well prove just as difficult to measure redshifts for 
the X-ray selected `quasar IIs' as has proved the case for their radio-loud
counterparts in the 7CRS (Willott, Rawlings \& Blundell 2000a):
nuclear obscuration means that quasar IIs can masquerade 
as extremely-red galaxies with only weak observable narrow 
emission lines in the near-IR, especially in the key 
$1.2 \leq z \leq 1.7$ redshift range.

One open question concerns the size of any Compton-thick population of AGN.
High-energy X-ray studies of local AGN (e.g.\ Maiolino et al.\ 1995)
suggest that, at least at low values of $L_{bol}$, most nuclei are obscured
by Compton-thick material, so the normalisation of the quasar luminosity
function in Fig.~2 is still strictly only a lower limit.
{\sc Chandra} follow-up of AGN selected at low radio frequencies might 
prove vital here since the large
radio emitting lobes should be unobscured, 
allowing direct measurement of the distribution in 
obscuring column densities for a randomly-oriented population of quasars.

Fig.~2 also illustrates the key remaining challenge 
for understanding the quasar population: why are there seemingly
$\sim100-$times more radio-quiet quasars than radio-loud 
quasars\footnotemark?
\footnotetext{
This 100:1 ratio of radio-quiet to radio-loud objects is somewhat higher than
most other estimates in the literature (e.g.\ Goldschmidt et al.\ 1999).
The main reasons for this are associated with the effects of
Doppler boosting. First, both the optical and (high-frequency) 
radio emissions from an intrinsically radio-loud object can be boosted 
such that more common (but intrinsically lower $L_{\rm bol}$) objects 
are preferentially promoted into an optically-selected sample of quasars
if they have powerful relativistic jets.
Second, low-power relativistic jets seem common amongst 
the radio-quiet population
(e.g.\ Miller, Rawlings \& Saunders 1993); when aligned
with the line-of-sight, these jets produce sufficient radio 
emission that they can be designated as `radio-loud'
despite the absence of an intrinsically powerful jet.
}
The most na\"{i}ve explanation -- that `radio-quiet'
phases of quasar activity last $\sim100-$times longer than
`radio-loud' phases -- is not really tenable.
Since jet-fed powerful radio sources persist alongside nuclear
optical/X-ray quasar activity for timescales
of $\sim10^{8}$ years (e.g. Blundell, Rawlings \& Willott 1999;
Blundell \& Rawlings 2000), then any such radio-quiet phase
would need to last $\sim10$ Gyr, leading to just one 
generation of quasars and a well-known problem explaining the
mass density of super-massive black holes in the 
local Universe (e.g.\ Richstone et al.\ 1998).
The cosmic
evolutions of the radio-loud and radio-quiet populations 
(Fig.~2) seem to be driven by a common cosmic cause, 
presumably $\sim10^{9}$-yr timescale variations in the trigger rate convolved 
with shorter-timescale ($\sim10^{8}$ yr) effects due to
finite quasar lifetimes. 

A more likely explanation for the large 
fraction of radio-quiet to radio-loud objects follows from the
invocation of a minimum black hole mass for the
development of powerful jets. Consider a toy model in which 
quasars divide cleanly between radio-quiet quasars and radio-loud quasars
at a critical black hole mass $M_{\rm BH} \sim 10^{9.5} ~ \rm M_{\odot}$:
because of the steepness of the mass function for black holes 
(e.g.\ Salucci et al.\ 1999), for every radio-loud
quasar there would be a factor $\gtsim 100$-times
more radio-quiet quasars with black hole masses in the 
range $8.5 \leq \log_{10} M_{\rm BH} < 9.5$. Thus, a spread 
of only $\sim1$ dex in the ratio of $L_{\rm bol} / M_{\rm BH}$
would cause the radio-quiet quasars to swamp radio-loud quasars even
at the highest values of $L_{\rm bol}$ (see also McLure et al.\ 1999).
There is some indirect support for the central idea here --
that, compared to radio-quiet quasars, the distribution in 
$M_{\rm BH}$ for radio-loud quasars has a high mean and a low variance --
from the normalisation and spread in the near-infrared Hubble 
diagram for radio galaxies (e.g.\ Eales et al.\ 1997). Although it seems 
hard to envisage a critical physical switch allowing
powerful jet activity only at 
$M_{\rm BH} \gtsim 10^{9.5} ~ \rm M_{\odot}$, it may
appear to be present because of the 
special formation history of the most massive black holes.
As explored by Wilson \& Corbett (1995), the major merger of two
massive dark haloes containing central black holes could 
result in the formation of a single super-massive spinning black hole
--- the preferred origin for powerful jets via the Blandford-Znajek
mechanism (1977) --- as well as the formation of the core of
a luminous elliptical galaxy via heating associated with 
the orbital decay of the binary black hole system (Faber et al.\ 1997).

Whatever the underlying physics, the evolutionary consequence of the
normalisations of the luminosity functions
of Fig.~2 and quasar lifetimes $\sim10^{8} \rm ~ yr$
is that every massive ($\gtsim L^{*}$) 
galaxy was plausibly a quasar at high 
redshift (e.g.\ Richstone et al.\ 1998), 
and that $\sim1$ in 100 of such galaxies
spent time as a powerful radio source. The present-day
(co-moving) space density of the sites of past Eddington-tuned
jet activity is given roughly by the density at $z \sim 2.5$
times the ratio of the duration of `the quasar epoch'
(see Sec.~4) to the duration of a radio source
outburst, or $\sim10^{-7} \times 4 \times 10^{9} / 10^{8} 
\approx 4 \times 10^{-6} ~ \rm Mpc^{-3}$ which turns out to be 
precisely the space density of rich clusters (Dalton et al.\ 1994).  
Plausibly, therefore, every rich cluster of galaxies was once the site 
of extreme radio activity.

\section{Cosmic evolution at $z > 2.5$: the redshift cut-off}
\label{sec:hiz}

\begin{figure}[!h]
\label{fig:entropy}
\plottwo{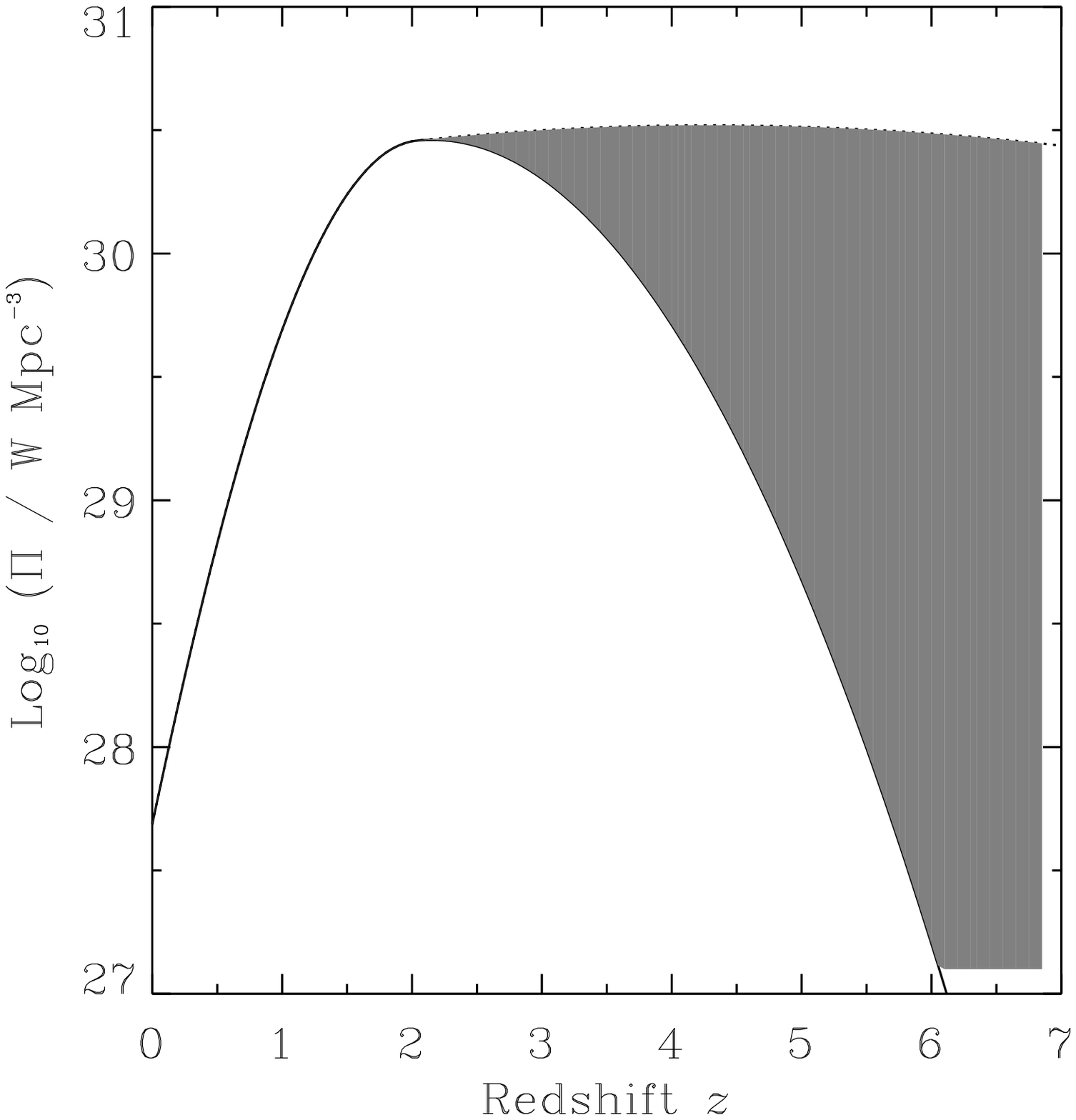}{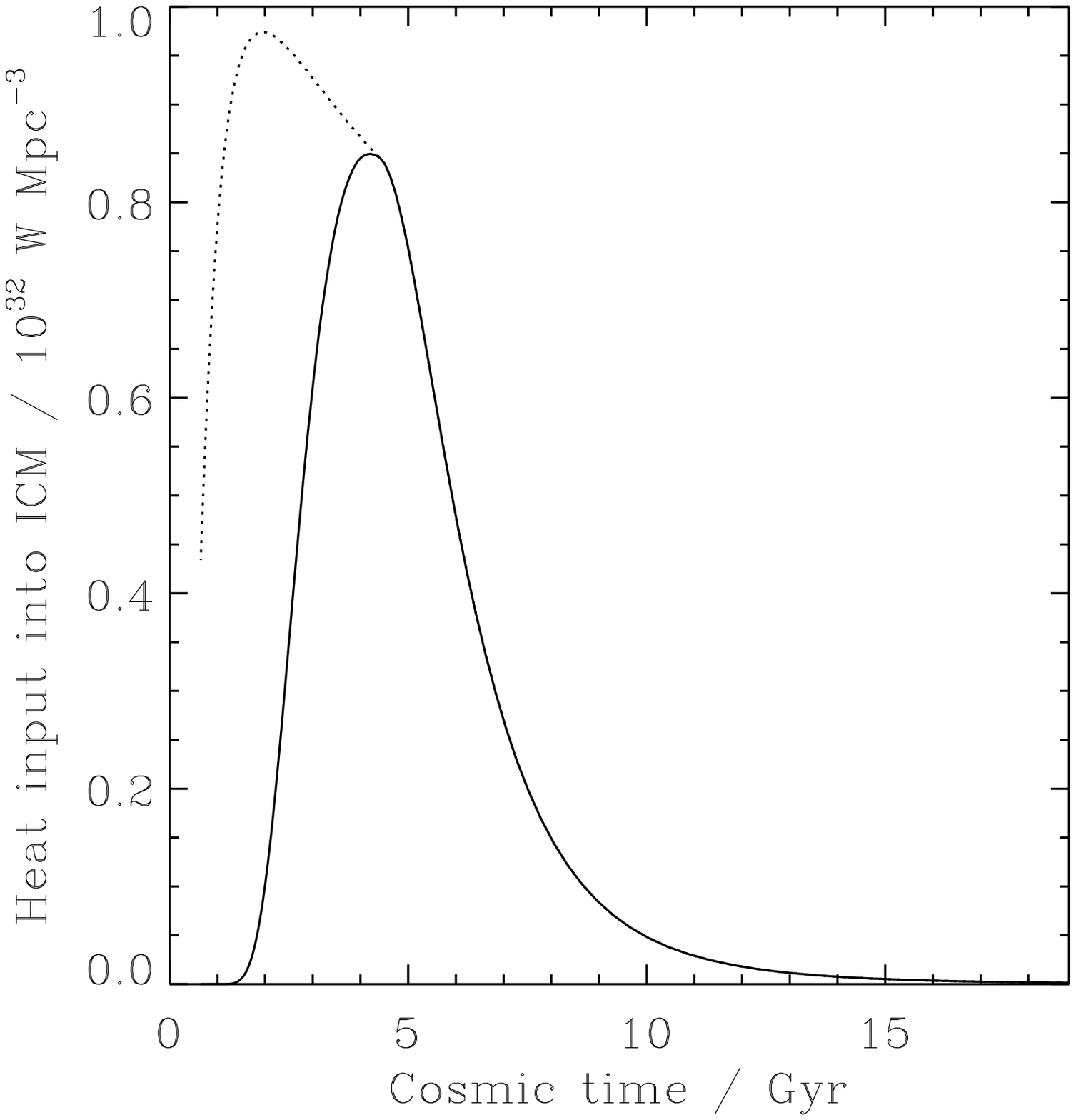}
\caption{({\bf Left}) Rough constraints on the high-redshift evolution of the
luminosity density $\Pi = \int \nu L_{\nu} 
[{\rm d} \rho / {\rm d} \log_{10} L_{\nu}] {\rm d}
(\log_{10} L_{\nu}) {\rm d} \nu$ of the
Eddington-tuned radio population (which dominates the 
total radio luminosity density at high $z$). These have been calculated from 
model C of Willott et al.\ (2000c), amended at high-redshift to  
match the $\sim90$ per cent confidence intervals on the high-$z$ 
decline in $\Pi$ inferred
(indirectly) from constraints on the decline in 
the apparent space density of the flat-spectrum
population (see Jarvis \& Rawlings 2000) --- direct constraints 
(e.g.\ Jarvis et al.\ 2000) are even looser! The limits taken for
the frequency integral were 10 MHz to 100 GHz in the rest-frame, 
and $\alpha=1$ was adopted. 
({\bf Right}) A rough calculation of the heat injected into the
intracluster medium (ICM) by Eddington-tuned radio sources per co-moving
Mpc$^{-3}$ as a function of cosmic time. The solid line 
shows the prediction made on the basis of the lower bound to 
the high-$z$ RLF in the left-hand figure, and the dotted line
represents the upper bound. This calculation makes the
following assumptions: (i) that the total jet power $Q$ per Mpc$^{3}$
can be obtained from the RLF using the transformation
$Q =  10^{16.2} \times L_{151}^{0.85}$ (Willott et al.\ 1999); and (ii)
that $50$ per cent of the jet power is dissipated as heat in the
bow-shock of the radio source (Rawlings \& Saunders 1991; 
Willott et al.\ 1999; Blundell \& Rawlings 2000).
}
\end{figure}

The results of Sec.~3 beg the question of constraints
on the evolution of the radio source population at high redshift, 
i.e.\ at $z > 2.5$. I will focus my attention on the 
Eddington-tuned population since, as is obvious from Fig.~1,
the lower-luminosity population is not detectable at high redshifts
in bright radio-selected samples. My contention here is that,
even though many radio-luminous objects have now been observed in the 
$z > 2.5$ Universe, constraints on their space density remain
very loose. Since these points are 
discussed in detail elsewhere (Jarvis \& Rawlings 2000;
Jarvis et al.\ 2000; Willott et al.\ 2000c), I will
simply summarise my view of the current situation.

\begin{itemize}

\item Direct studies of the steep-spectrum population at
$z > 2.5$ based on redshift surveys of complete samples
(Willott et al.\ 2000c) do not allow us to discriminate between models in 
which the $\Upsilon$ of radio sources declines at high redshift
from those in which it stays constant.
This is a subtly different conclusion to that reached by
Dunlop \& Peacock (1990), and the possible reasons for this difference
are discussed by Willott et al.\ (2000c).

\item Through finding steep-spectrum radio galaxies at $z \sim 4-5$
(e.g.\ Rawlings et al.\ 1996),
direct studies of the $z > 2.5$ population using `filtered'\footnotemark
samples suggest roughly constant $\Upsilon$ from $z \sim 2.5$ to $z \sim 5$,
although gradual declines or even inclines 
cannot be ruled out (Jarvis et al.\ 2000).

\footnotetext{
A filtered sample is one in which optical spectroscopic follow-up is confined
to a small fraction of the radio sources in a flux-density limited sample, 
where the filters used by current surveys are a (steep) spectral index
criterion and/or a (small) angular size criterion (e.g.\ Blundell et al.\
1998).
}

\item Indirect\footnotemark evidence for a high-$z$ cut-off in $\Upsilon$
from the study of luminous flat-spectrum radio quasars is more equivocal than 
received wisdom might suggest. 
Gradual declines in the space density of the most luminous flat-spectrum
quasars, by a factor $\sim4$ between $z \sim 2.5$ and $z \sim 5$ 
(Dunlop \& Peacock 1990), are favoured by the data but the 
observable co-moving volume is too small to make definitive statements:
both constant space density models and abrupt cut-offs, like those
suggested by Shaver et al.\ (1996), are ruled out only at the 
$\approx 2 \sigma$ level (Jarvis \& Rawlings 2000).

\end{itemize}

\footnotetext{
Eddington-tuned jets pointed within a few degrees of the
line-of-sight are Doppler-boosted to produce flat-spectrum
radio quasars; these are thought to be the dominant
flat-spectrum population in high-frequency selected 
radio samples [although see Jarvis \& Rawlings (2000) for a discussion
of the importance of GHz-Peaked sources]. The apparent space density
of these flat-spectrum sources gives an indirect measure of the 
space density of the parent steep-spectrum population close to, but above,
the break in the RLF (Jarvis \& Rawlings 2000).
}

This state of ignorance concerning the high-$z$ evolution of radio 
sources is illustrated in Fig.~3.
It is worth noting that it is not clear that any other 
published measures of the high-$z$ evolution of quasars are any 
better constrained, largely because of the difficulty of correcting 
for obscuration in optical/soft X-ray selected samples.
In fact the abrupt cut-off found by 
optical quasar surveys (e.g.\ Warren, Hewett \& Osmer 1994) 
and the roughly constant
space densities found by X-ray surveys (e.g.\ Miyaji et al.\ 2000)
are together suggestive of obscuration: as a spectrum becomes increasingly
redshifted, attenuation of 
optical light becomes more effective due to the increased efficiency of
dust obscuration at UV wavelengths,  
whereas the opposite is true for the effects of redshifting on the
opacity at X-ray wavelengths (e.g.\ Wilman \& Fabian 1999).

A large number of programmes are compromised by ignorance
concerning the redshift cut-off,
including basic questions like: (i) whether or not the onset of
quasar activity re-ionized the Universe (e.g.\ Haiman \& Loeb 1997);
(ii) understanding the nature of the sub-mm
source counts (e.g. Almaini, Lawrence \& Boyle 1999); and (iii) 
using the density of high-$z$ objects to constrain the
power spectrum of density fluctuations on $\sim$Mpc 
scales (e.g.\ Efstathiou \& Rees 1988).

I concentrate here, however, on 
an important physical parameter connected directly to 
the high-$z$ evolution in radio luminosity density ---
the contribution of Eddington-tuned radio sources to the 
entropy budget of the Universe. The results of a rough calculation are
shown in Fig.~3. The key point is that 
although the radio source phenomenon is $\sim100-$times rarer than 
the quasar phenomenon, the relativistic jets feeding
radio sources have bulk powers which are of the 
same order as the Eddington luminosity $L_{\rm Edd}$
of a $\sim10^{9} ~ \rm M_{\odot}$ black hole
(Rawlings \& Saunders 1991; Willott et al.\ 1999; 
Blundell \& Rawlings 2000), and 
$\gtsim 100-$times more powerful than the relativistic jets in
luminous radio-quiet quasars (Miller, Rawlings \& Saunders 1993). So,
provided that the power in quasar winds is $\ltsim 0.01~L_{\rm Ledd}$
(Silk \& Rees 1998) and provided one can neglect photoionisation
heating (Valageas \& Silk 2000), it seems plausible
that more power is available to heat the IGM as the result of 
radio source induced shocks, than is yielded by the
more numerous radio-quiet quasar population.
This heat, or entropy, input into the IGM will be concentrated in the
clusters of galaxies which at one time hosted powerful
radio sources. For the cut-off model in 
Fig.~3, the duration of the 
quasar epoch is $\sim4 ~ \rm Gyr ~ (\sim 10^{17} s)$ and  
a crude estimate of the heat input into each rich cluster 
by Eddington-tuned radio sources is
$\sim 10^{32} \times 10^{17} / 4 \times 10^{-6} 
\approx 2.5 \times 10^{54} ~ \rm J$. This is a reasonable fraction
of the total thermal energy ($\sim5 \times 10^{55} ~ \rm J$)
in a rich ($\sim10^{14} ~ \rm M_{\odot}$
in baryons) cluster.
For the no-cutoff model of Fig.~3, 
the quasar epoch lasts longer, and the heat input 
approaches $\sim10$ per cent of the total thermal energy in rich clusters.
Understanding Fig.~3 is therefore likely to be
central to any understanding of the entropy budget of the Universe.
At earlier cosmic times the heat input by radio sources was probably
sufficient to overcome the gravitational binding energy of gas 
in the relatively shallow potential wells of
forming clusters; this would provide an attractive alternative to
supernova- and quasar-driven mechanisms invoked to 
explain the `excess' entropy (over and above that due to
shock heating during cluster collapse)
identified by studies of the properties
of rich clusters (e.g.\ Kaiser 1991).
In essence I am suggesting a feedback mechanism
similar to the quasar-driven wind model
invoked for galaxy formation by Silk \& Rees (1998)
but applicable to more massive systems, i.e.\ rich clusters, and to
more powerful outflows, i.e.\ radio source driven bow shocks.

\section{The sub-mm view of the cosmic evolution of radio sources}
\label{sec:submm}

\begin{figure}[!h]
\label{fig:submm}
\plottwo{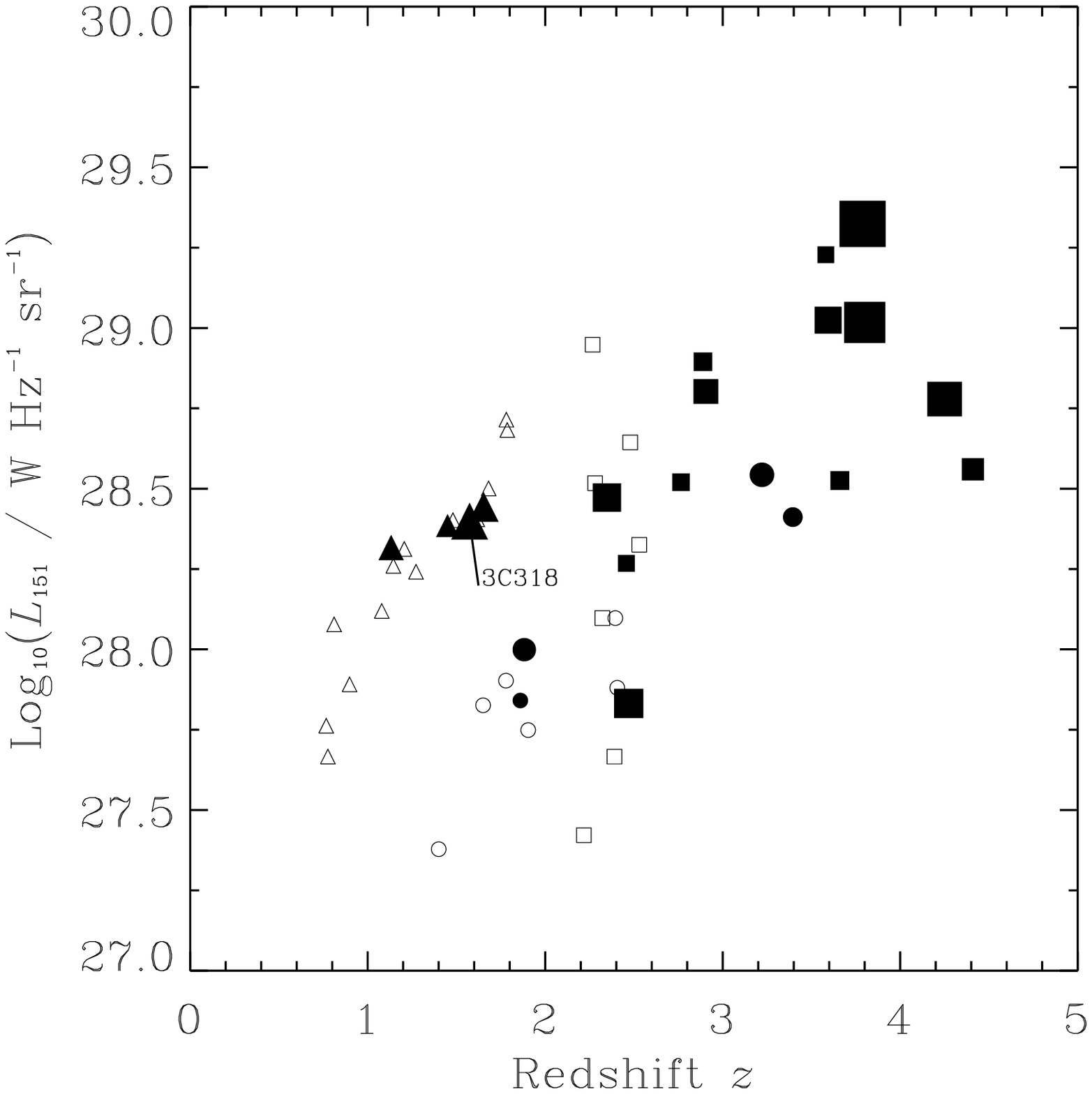}{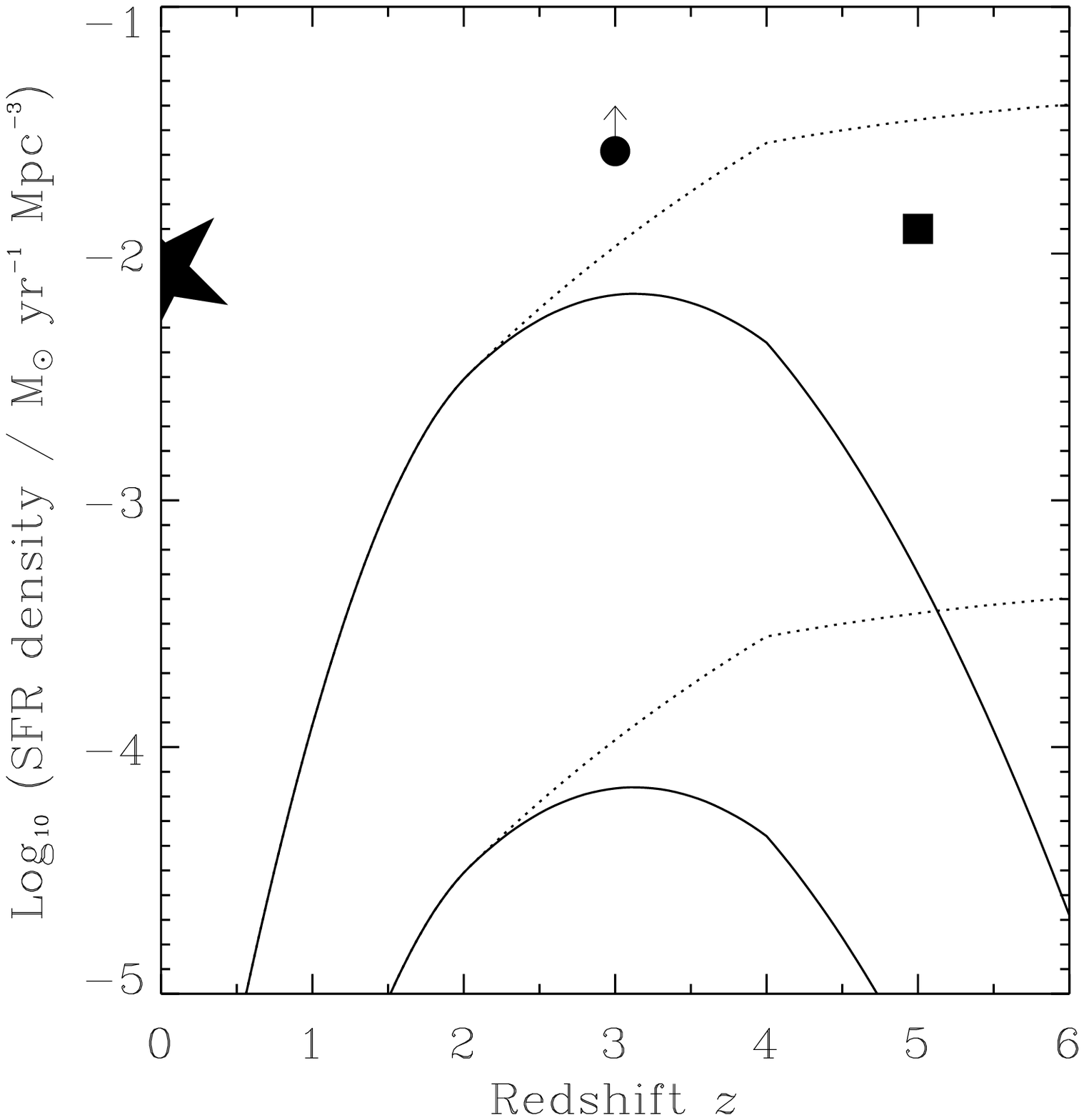}
\caption{({\bf Left})
$L_{151}$ versus $z$ plane
for 47 radio galaxies [one at $\log_{10} (L_{151}) = 26$, $z =1.43$ is
off the plot], and one radio quasar 3C318, observed at 
$350 ~ \rm GHz$ with SCUBA (triangles -- 3CRR objects,
circles -- 6CE objects, squares -- objects from 
other surveys); filled symbols represent
detected objects, and open symbols those with upper limits.
The sizes of the symbols are scaled from the $2 \sigma$
350 GHz flux density $S_{350}$ limit ($S_{350} \approx 2 ~ \rm mJy$)
by the factor
$[1 + \ln (S_{350} / 2)]$.
({\bf Right}) Illustration of the fraction of global star-formation
associated directly with the quasar population.
The star-formation associated with radio-loud objects
is shown by the two lower lines adopting either the lower bound to 
the high-$z$ RLF (solid line, the cut-off model of Fig.~3) or
the upper bound (dotted line, no cut-off model). This has been
calculated by assigning a star formation rate (SFR) of $10 (1+z)^3 
~ \rm M_{\odot} ~ yr^{-1}$ (saturating at $z=4$) to each Eddington-tuned
radio source, without any correlations between $L_{151}$ and SFR.
To account for the radio-quiet population the upper lines 
scale these functions by a factor of 100 (see Sec.~3).
The symbols represent the following: star, local star formation
rate from the H$\alpha$ survey of Gallego et al.\ (1995); circle,
SCUBA HDF survey (Hughes et al.\ 1998), given by Hughes et al.\
as a lower limit but likely to be within a factor of a few of the correct
value given the brightness of the sub-mm background; square, 
global SFR corresponding to one of the SCUBA HDF sources lying at
$4 \leq z < 6$.
}
\end{figure}

The
strong cosmic evolution of the Eddington-tuned quasar population 
discussed in Sec.~3 is most likely the 
simple and direct consequence of the standard hierarchical model
for galaxy formation in which major mergers of galaxy-sized
sub-clumps are more prevalent in the younger Universe.
A prediction of this hypothesis is that the Eddington-tuned radio population
is associated with sites of intense star formation since, at high $z$,
a large fraction of the baryons in the sub-clumps are likely to be in
the form of gas rather than stars 
(e.g.\ Storrie-Lombardi, McMahon \& Irwin 1996). 
The observed properties of
high-$z$ radio sources should show signs of this.
Although UV and optical observations have now
provided at least one compelling example of a huge starburst 
associated with a high-$z$ radio galaxy (Dey et al.\ 1997), 
it has proved impossible to prove the ubiquity of the starburst
phenomenon from UV/optical/near-IR, or indeed radio, observations
because any starburst signature can be so easily swamped by emission intimately
associated with the active nucleus, e.g.\ in the case of UV observations, 
scattered light from the quasar nucleus. A far cleaner
diagnostic of large rates of star formation in 
radio galaxies is a huge rest-frame far-infrared luminosity\footnotemark, 
requiring a survey of high-$z$ radio galaxies in the sub-mm waveband. 
The advent of the SCUBA camera on the JCMT has recently made 
such a survey possible. Full details of this work are published elsewhere 
(Archibald et al.\ 2000), so I concentrate here on 
the major results and their implications. 

\footnotetext{
Photometric measurements on the Rayleigh-Jeans side of 
a thermal spectrum can be used to confirm that the emission is due to dust
(which for radio galaxies it almost always is)
and provide good estimates of dust mass which, in well-studied 
objects, correlates closely with gas mass 
(e.g.\ Hughes, Dunlop \& Rawlings 1997). Since it is 
often difficult to prove beyond doubt that the dust is heated by stars rather
than an active nucleus (e.g.\ Willott, Rawlings \& Jarvis 2000e), the 
correlation between dust/gas mass and star formation rate
is more uncertain.
}

Fig.~4 shows how the chances of making a sub-mm detection
of a radio galaxy varies across the $L_{151}$--$z$ plane.
Considering $z \gtsim 3$ and $\log_{10} (L_{151}) \gtsim 28.5$, we see that
all the radio galaxies are 
detected by SCUBA, whereas at lower redshifts and $\log_{10}(L_{151})$
the fraction of detections falls abruptly to $\ltsim 20$ per cent.
Archibald et al.\ make a detailed study of the underlying correlations
between rest-frame far-IR luminosity $L_{\rm FIR}$, $L_{151}$ and $z$.
They conclude that the major effect at work is an 
$\sim(1+z)^{\sim 3}$ rise in $L_{\rm FIR}$ for radio galaxies 
out to $z \gtsim 4$. The most straightforward 
interpretation of this result is that 
the characteristic star formation rate (SFR) of luminous radio galaxies rises
from $\ltsim 10 ~ \rm M_{\odot} ~yr^{-1}$ at $z = 0$ to $\gtsim 1000 ~
\rm M_{\odot} ~yr^{-1}$ at $z \sim 4$, 
an interpretation which seems to fit in naturally with the
arguments made in Sec.~3.
If the strong cosmic evolution 
in $\Upsilon$ for the Eddington-tuned radio population is indeed driven by
systematic changes in the trigger rate with $z$ then the  most viable trigger
mechanisms, e.g. the formation of a 
rapidly spinning super-massive black hole in
a major merger event, are likely to be far more prevalent at high $z$;
provided, as seems likely,
the progenitor galaxies are gas rich, intense associated
starbursts seem inevitable. 

In Fig.4 I take these arguments to their logical
conclusion by deriving rough estimates of the
star formation in the Universe associated directly with 
AGN. It seems that, as noted 
by Almaini et al.\ (1999), a large fraction ($\gtsim 10$ per cent)
of the star formation in the high-$z$ Universe is probably
occurring directly alongside quasar activity in the cores of forming 
galaxies (see also Boyle \& Terlovich 1998). Indeed, if we hypothesise
that galaxies form in a series of short (less than or of the order of a
quasar lifetime\footnotemark), intense ($\gtsim 100 ~ \rm M_{\odot}
~ yr^{-1}$) starbursts across the entire quasar epoch 
(a process which perhaps culminates in the quasar phase) 
then the important star formation at high redshift goes hand-in-hand
with the formation of black holes, and the epoch of major quasar
activity. Of course if Compton-thick objects turn out to be the dominant
AGN population at high-$z$, then the link between AGN and star formation
activity may prove closer still.

\footnotetext{
One of many caveats to my interpretation of 
Fig.~4 concerns the question of the 
relative timescales of AGN and star-formation activity.
Blundell \& Rawlings (1999) have pointed out that selection effects
ensure that the known $z >3$ radio galaxies, e.g. those in the SCUBA survey of
Archibald et al.\ (2000), are viewed within $10^{7}$ yr
of the jet-triggering event. If the characteristic timescale of the
starburst activity  is shorter than the $\sim10^{8} ~ \rm $yr
timescale of AGN activity, then the `youth-redshift
degeneracy' (Blundell \& Rawlings 1999) may be an important
contributor to the observed `cosmic evolution' in the 
SFR of powerful radio galaxies (Fig.~4). 
It is interesting in this regard that
one of the youngest 3CRR radio sources at $z \sim 1.5$, 3C318,
is a SCUBA-detected quasar (Willott et al.\ 2000e). 
}

The results of this section concerning the history of
star formation, and of Sec.~4 concerning the 
global entropy budget, should provide food for thought for the
growing band of researchers using semi-analytic models to explain
galaxy formation and evolution (e.g.\ Somerville \& 
Primack 1999). At present these models take no account of feedback 
between quasars, their jets and 
their host galaxies. For observers of
high-$z$ AGN it is obviously vital
to tie down huge and basic uncertainties like the
evolutionary behaviour of quasars at $z > 2.5$.

\section{Future prospects: bigger redshift surveys and better modelling}
\label{sec:future}

In previous sections we have seen how
cosmological studies using low-frequency selected
radio sources are compromised by the limitations of the 
current generation of redshift surveys.
With these considerations partly in mind Gary Hill ({\sc Tex}as) 
and I ({\sc Ox}ford) have begun a programme of {\sc TexOx} redshift
surveys. One aspect of this programme --- 
the {\sc TexOx-1000} (or {\sc TOOT}) survey --- is to measure $\sim1000$ 
redshifts for a complete sample flux-limited 
at $S_{151} = 100 ~ \rm mJy$. The optical/radio/near-IR
imaging part of this survey is now almost complete and spectroscopic 
follow-up began in earnest in January 2000. 
This programme aims to exploit the 
complementary capabilities of UK facilities like the ISIS 
spectrometer on the WHT -- the instrumental back-bone of the
6CE and 7CRS programmes -- and the LRS spectrometer on the
new 9.2-m Hobby-Eberly Telescope.
We are already about a quarter-way through the required spectroscopic
follow-up, and we anticipate completion of the {\sc TOOT} survey in mid 2002.
As illustrated by Fig.~1, the {\sc TOOT} survey will be the
first to directly measure the space density of
$\log_{10}(L_{151}) \gtsim 26.5$ objects, and hence the 
luminosity density of the steep-spectrum population,
at $z \sim 2.5$.

Still larger redshift surveys may be needed to settle the 
question of evolution at $z \gg 2.5$. A major difficulty of making 
such surveys is the lack of any 
multiplex capability at relatively bright 
flux-density limits. At the $S_{151} \sim100 ~ \rm mJy$ limit of the 
{\sc TOOT} survey, the average separation of sources on the sky is $\sim 20$ 
arcmin so
there is still no effective alternative to
object-by-object spectroscopic follow-up, 
particularly as the largest field-of-view 
multi-object systems tend to compromise one or more of sensitivity and
spectral coverage.
In the medium-term future, progress can probably best be made
by refining photometric-redshift techniques for the radio source
population, although the requirement for wide-field
capability at near-IR wavelengths 
means that substantial progress will probably require
new initiatives like the planned surveys with {\sc VISTA}
(see Jarvis et al.\ 2000). In the long-term future, the hope is that 
redshifts will come for free as facilities like
the Square Kilometre Array push radio sensitivities to the 
level where surveys can routinely detect HI and CO features 
superimposed on the radio continuum.

As we await the results of bigger redshifts surveys, 
a high priority is to build better models for the cosmic evolution of the 
radio source population. This splits into two categories of problem.
First, we need to develop robust models for the physical evolution of 
radio sources: i.e.\ once a source of a given jet power has been triggered,
how does its radio properties (e.g. luminosity, size and spectral index)
develop as a function of time, and as a function of the radio source
environment? Second, we need to understand systematic changes
in radio source environments with $z$, both because of their influence 
on radio properties, but more importantly so that we can understand how
jet trigger rates reflect the build up
of structure in the Universe. In the first area, the
hiatus following the pioneering work of Scheuer (1974)
is now well and truly over (e.g.\ Kaiser,
Dennett-Thorpe \& Alexander 1998; Blundell et al.\ 1999).
In the second area, semi-analytic models could provide the 
way forward, so these must now be developed to 
incorporate the feedback between AGN, their jets and their
environment.

\section{Acknowledgments}

The Pune meeting was excellent and I would like to 
thank all those involved in organising it. Much of the work
discussed in this paper stems from collaborative research involving 
the following: Elese Archibald,
Katherine Blundell, Steve Croft, Jim Dunlop, Steve Eales, Pamela Gay,
Gary Hill, Dave Hughes, Rob Ivison, Matt Jarvis,
Mark Lacy and last, but certainly not least, Chris Willott.

\end{document}